\documentclass{aa}
\usepackage{pjournal}
\usepackage{graphicx}
\usepackage{epstopdf}
\usepackage{txfonts}
\usepackage{subeqnarray}
\usepackage{natbib} 
\usepackage{version}
\usepackage{rotating}
\usepackage{aurelie}

\begin{document}

\title{Narrow band selected high redshift galaxy candidates contaminated by lower redshift O[III] ultrastrong emitter line galaxies
\thanks{Based on observations made with ESO telescopes at the La Silla Paranal Observatory under programmes ID 385.A-1025(A) and 181.A-0485.}}
\author{Aur\'elie P\'enin \inst{1,2}
\and Jean-Gabriel Cuby \inst{1}
\and Benjamin Cl\'ement \inst{3,6}
\and Pascale Hibon\inst{5}
\and Jean-Paul Kneib\inst{1,4}
\and Paolo Cassata\inst{1}
\and Olivier Ilbert\inst{1}
}
\institute{Aix Marseille Universit\'e, CNRS, LAM (Laboratoire d'Astrophysique de Marseille) UMR 7326, 13388, Marseille, France 
\and Astrophysics and Cosmology Research Unit, School of Mathematical Sciences, University of KwaZulu-Natal, Durban 4041, South Africa 
\and Steward Observatory, University of Arizona, 933 North Cherry Avenue, Tucson, AZ 85721, USA
\and Laboratoire d’astrophysique, Ecole Polytechnique F\'ed\'erale de Lausanne (EPFL), Observatoire de Sauverny, CH-1290 Versoix, Switzerland
\and Gemini Observatory, Casilla 603, La Serena, Chile
\and CRAL, Observatoire de Lyon, Université Lyon 1, 9 Avenue Ch. André, 69561 Saint Genis Laval Cedex, France
}
\abstract{Lyman Break Galaxies (LBG) and Narrow Band (NB) surveys have been successful at detecting large samples of high-redshift galaxies. Both methods are subject to contamination from low-redshift interlopers.}
%
{In this paper, our aim is to investigate the nature of low-redshift interlopers in NB Lyman-$\alpha$ emitters (LAE) searches.}
%
{From previous HAWK-I NB imaging at $z \sim 7.7$ we identify three objects that would have been selected as high-redshift LAEs had our optical data been one magnitude shallower (but still one to two magnitudes fainter than the near infrared data). We follow-up these objects in spectroscopy with XSHOOTER at the VLT.}
%
{Despite low quality data due to bad weather conditions, for each of the three objects we identify one, and only one emission line, in the spectra of the objects, that we identify as the O[III]5007$\AA$ line. This result combined to spectral energy density fitting and tests based on line ratios of several populations of galaxies we infer that the 3 objects are ultrastrong line emitters at redshifts $\sim 1.1$.}
%
{From this work and the literature we remark that the O[III] line appears to be a common source of contamination in high-redshift LBG and LAE samples and we suggest that efforts be put to characterize with high accuracy the O[III] luminosity function out to redshift $\sim 3$ or higher.}

\titlerunning{The role of the O[III] line in high-$z$ samples}
\maketitle 

\section{Introduction}

The study of the highest redshift galaxies and quasars at $z >7$ provides insights into the early stages of galaxy formation and evolution and on the ionization state of the Inter Galactic Medium (IGM). Much progress has been achieved over the past decade in assembling samples of high redshift galaxies from essentially two techniques: the Lyman Break technique selecting Lyman Break Galaxies (LBG) and the Narrow Band (NB) imaging technique targeting Ly$\alpha$ emitters (LAEs).\\
The Lyman Break method looks at the Lyman continuum break in the UV spectra of galaxies. Pioneered by \citet{2003ApJ...592..728S} at redshift $\sim$ 3, this technique has been progressively applied to higher redshifts. As redshift increases, the Ly$\alpha$ forest becomes so dense that the continuum blueward of the Ly$\alpha$ line is almost entirely suppressed, becoming in practice the signature used to select high-redshift galaxies with the LBG technique. At redshift 7, the break occurs around 1 $\mu$m and galaxies can only be observed in the near infrared.\\
In several studies performed after the installation of the WFC3 camera on board HST in 2009, the Lyman Break technique has been used to search for galaxies up to redshifts 10 or higher. Samples of hundreds of galaxies at redshifts 7 to 9 have been assembled, and while their spectroscopic confirmation remains an extremely difficult endeavor due to their faintness, the high redshift nature of these samples is extremely robust and indisputable, with estimated contamination rates of the order of 10 to 20\%. \citet{2014arXiv1403.4295B} provides a recent and comprehensive compilation of $z \sim 4$ to $z \sim 10$ LBG galaxy samples from various HST datasets. The $z\sim 7$ and $z \sim 8$ samples have respectively approximately 600 and 200 objects, with an estimated contamination rate of about 10\%, much lower than the contamination rate estimated on earlier samples \citep{2011ApJ...727L..39T}.\\
Searching high-redshift Ly$\alpha$ emitters from NB imaging relies on combining a flux excess in the NB filter with a break at the wavelength of the NB filter that corresponds to the break below the Ly$\alpha$ line mentioned above. At least one broad band filter overlapping with, or redder than, the NB filter is required to select line emitting objects, and additional redder broad band filters may provide supplemental information on the colors of the objects. The narrow band filters are usually centered on regions of low OH emission from the sky for optimum sensitivity, therefore leading to discrete redshift intervals, namely 5.7, 6.5, 7.3, 7.7 and 8.8. This technique has been extremely successful at finding $z \sim 6.5$ LAES \citep{2004AJ....127..563H,2006ApJ...648....7K,2010ApJ...723..869O,2010ApJ...725..394H,2011MNRAS.412.2579N,2011A&A...525A.143C}. These samples have been robustly confirmed in spectroscopy, with estimated contamination rates of the order of or below 30\%. The technique has been used into the near IR to search for even higher redshifts, see e.g. \citet{2007A&A...461..911C,2008MNRAS.384.1039W} for pioneering searches at redshifts 8.8 with small format near IR detectors. With the advent of large mosaics of near IR arrays, more recent searches have been carried out at redshifts 7.3, 7.7 and 8.8 \citep{2010ApJ...721.1853T,2010A&A...515A..97H,2011ApJ...741..101H,2012ApJ...745..122K,2012A&A...538A..66C,2012ApJ...752..114S,2013A&A...560A..94M,2014MNRAS.440.2375M,2014ApJ...797...16K}. Thus far, only one of the LAE candidates from these searches have been confirmed spectroscopically at $z \sim 7.3$. It was found with the Suprime-Cam CCD camera equipped with a special NB filter at the Subaru telescope \citep{2014ApJ...797...16K}.\\
Both the LBG and the NB observations have now convincingly pointed to a decline of the Ly$\alpha$ radiation at redshifts $> 7$, possibly due to an increasing fraction of neutral hydrogen from $z \sim 6$ when the Universe was completely re-ionized. Evidence for this attenuation of Ly$\alpha$ radiation comes from the strong decline of the comoving density of LAEs inferred from NB searches, as well as from the declining fraction beyond $z \sim 6.5$ of Ly$\alpha$ emitters among Lyman Break Galaxies \citep{2012ApJ...744..179S,2014arXiv1403.5466P,2014arXiv1405.4869T}. From the hundreds of LBGs selected at redshits $>7$, less than a handful have been spectroscopically confirmed \citep{2011ApJ...730L..35V,2013Natur.502..524F}.\\
Throughout these intensive searches for high redshift galaxies, several cases of contamination by low redshift interlopers or spurious sources have been reported. Contamination of high redshift LBG and LAE photometric samples can affect the constraints derived from these samples on the properties of high-redshift galaxies or on the neutral fraction of the Universe during the re-ionization epoch. It is expected that contamination increases with redshift, when the objects are fainter and less constrained from their photometry since the object may be detected in one or two bands only. This is exemplified by the spectroscopic identification at a redshift of 2.08 of an LBG candidate at $z \sim 11$ \citep{2012MNRAS.425L..19H} and the likely identification at a redshift of 2.19 of another LBG candidate at $z \sim 12$ \citep{2013ApJ...765L...2B,2013ApJ...773L..14C}. In an independent parallel survey with HST/WFC3, \citet{2011ApJ...743..121A} show that extreme emission line galaxies at $z\sim1-2$ can mimic the broad band colors of $z \sim 8$ LBGs, the high equivalent width lines being mistaken for the break at the Ly$\alpha$ line. Similarly, ultrastrong emission line galaxies can contaminate LAE samples by mimicking a color break on the blue side of the NB filter. Ultrastrong emission line galaxies selected from optical NB data are studied e.g. in \citet{2007ApJ...668..853K}, while \citet{2014MNRAS.440.2375M} and \citet{kochia} are two recent studies of NB-selected objects in the near infrared.\\
In this paper, we investigate the nature of possible contaminants of high-redshift LAE candidates, using our own NB data at 1.06~$\mu$m. To do so, we use the LAE samples from one of our previous studies \citep{2012A&A...538A..66C} in which we have selected high equivalent width line emitting objects at low redshift that would have been selected as LAE candidates at $z\sim7.7$ if the optical observations had been one magnitude shallower
From VLT-XSHOOTER  spectroscopy we investigate the nature of these objects.\\
We use AB magnitudes throughout this paper. We assume a flat $\Lambda$CDM model with $\Omega_\mathrm{m}$ = 0.30.\\ 
We describe the observations in Sect. \ref{par:data} and investigate the nature of these objects in Sect. \ref{par:which_obj}. Finally, we discuss the implications of our results for high redshift surveys in Sect. \ref{par:disc}.

\section{Data}\label{par:data}

\subsection{Photometry}

\begin{figure*}\centering
    \includegraphics[scale=0.40]{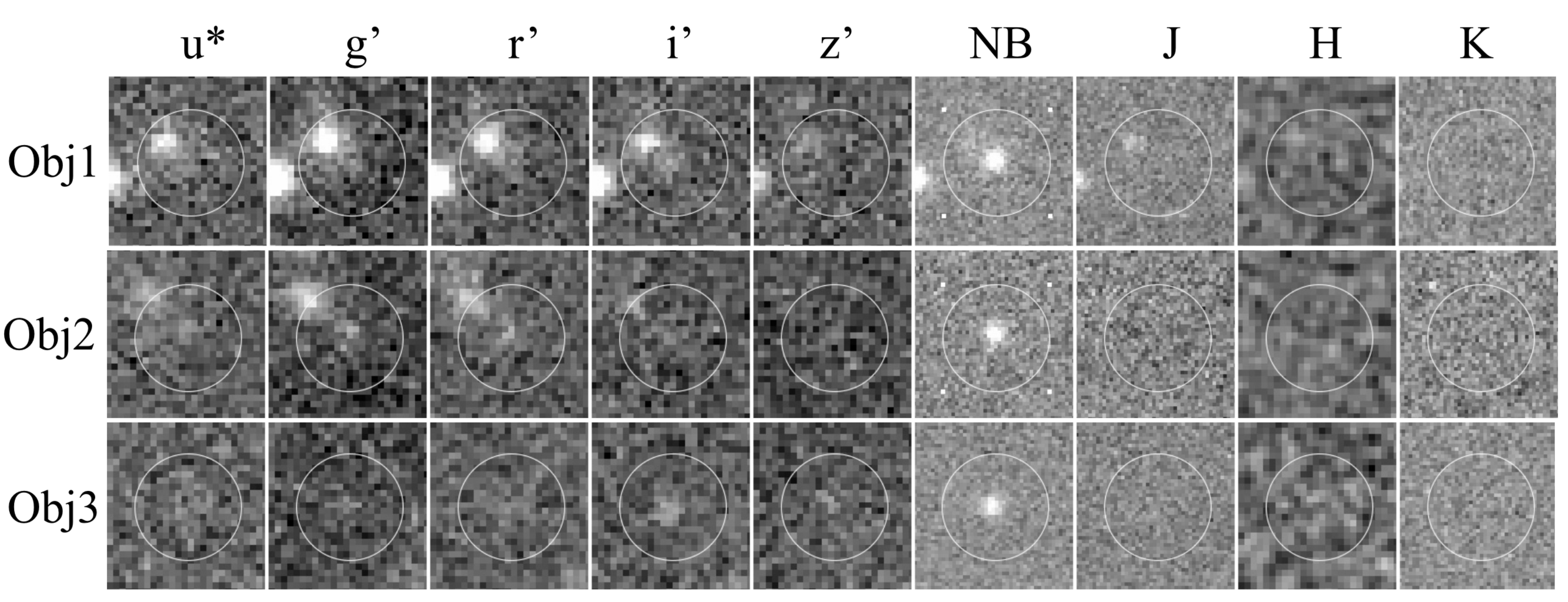}
  \caption{Thumbnail images of the three objects. The size of the thumbnails is 2.65$\times$ 2.65 arcsec$^2$.} 
  \label{fig:thumb_all}
\end{figure*}

\citet{2012A&A...538A..66C} present the results of a deep NB search of LAEs at 1.06 $\mu$m (corresponding to Ly$\alpha$ at $z \sim 7.7$) performed at the VLT with the HAWK-I instrument. This study did not identify $z = 7.7$ LAEs, allowing us to infer strong constraints on the evolution of the Ly$\alpha$ luminosity function from redshift 6.5. The NB survey reached a 3$\sigma$ aperture corrected limiting magnitude of 26.65, while the CFHTLS optical data have limiting magnitudes of 27.4, 28.2, 28.0, 27.4, and 26.6 in the u$^\star$, g', r', i', z' respectively.\\
The \nb~data were acquired over two epochs separated by one year to prevent the selection of transient object that would be detected in a one-epoch stack and not in the other. 
To select high redshift candidates, \citet{2012A&A...538A..66C} applied the following criteria:
\begin{enumerate}
\item \nb~$\geq 5\sigma\wedge$\nb$_\mathrm{epoch1}\geq 2\sigma\wedge$\nb$_\mathrm{epoch2}\geq 2\sigma$
\item No detection above the 2$\sigma$ level in any of the visible broad band filter
\item $2\leq Y - NB1060\leq 2.7$
\item \nb$-J\leq 0$ with 1$\sigma$ significance
\end{enumerate}
While Criteria 4 prevents the selection of T-dwarfs stars which usually have \nb -$J\geq 0$, Criteria 2 corresponds to the following colour criteria :
\begin{itemize}
\item $u^{\star}_{2\sigma}$ - \nb $\geq$ 1.7
\item $g^{\prime}_{2\sigma}$ - \nb $\geq$ 2.5
\item $r^{\prime}_{2\sigma}$ - \nb $\geq$ 2.3
\item $i^{\prime}_{2\sigma}$ - \nb $\geq$ 1.8
\item $z^{\prime}_{2\sigma}$ - \nb $\geq$ 0.9
\end{itemize}
In the process of selecting high-redshift candidates, seven peculiar objects were identified in the CFHTLS-D4 field. They are moderatly bright in the \nb~filter with AB$\leq$25.3, with SNR $\geq 5\sigma$ while they have  \nb - J $<$-1.5 and, in particular, z - \nb$>$1.8, typical of high redshift LAEs. Nevertheless, they are detected in optical broadband filters with magnitudes $\geq 27$ in the $u^\star$, $g^\prime$, $r^\prime$, $i^\prime$ filters and $\geq 26.5$ in the $z^\prime$. Clearly, these strong line emitting objects would have been selected as genuine LAE candidates with slightly shallower optical data. We observed the three brightest objects in spectroscopy with XSHOOTER to elucidate their nature. The photometric properties of these three objects are reported in Table \ref{tab:mag} and thumbnail images in all photometric bands are shown in Fig.  \ref{fig:thumb_all}.

\begin{table*}\centering 
\begin{tabular}{cccccc}
\hline\hline
instrument         &filter       &central wavelength (nm)    &  Obj1                    &Obj2                                   & Obj3 \\                        
\hline 
WIRCam             &Ks           &2146                                  &$>$24.70                &$>24.70$                         &$>$24.70 \\                   
WIRCam            &H            &1631                                  &$>$24.70                &$>$24.70                         &$>24.70$    \\               
HAWK-I                & J             &1258                                  &$>$26.53                &$>$26.53                         &$>$26.53       \\            
HAWK-I                &\nb         &1062                                  &24.3$\pm$0.1         &24.30$\pm$0.06              &24.52$\pm$0.05       \\
Megacam           &z'             &890                                    &$>$26.6                 &26.59$\pm$0.40              &26.43$\pm$0.31       \\
Megacam           &i'              &770                                    &27.1$\pm$0.2       &$>$27.43                          &27.02$\pm$0.19       \\
Megacam           &r'              &625                                    &27.3$\pm$0.2       &27.41$\pm$0.21              &$>$28.0                    \\
Megacam           &g'             &486                                    &27.3$\pm$0.2        &27.44$\pm$0.19             &27.89$\pm$0.27      \\
Megacam           &u$^{\star}$            &378                                    &27.1$\pm$0.3        &$>$27.37                        &$>27.37$                  \\
\hline 
\end{tabular}
\caption{Photometric data of the three NB selected objects studied in this paper. The significance of the limiting magnitudes is 3$\sigma$.}
\label{tab:mag}
\end{table*}

\subsection{X-SHOOTER observations}
The objects were observed with the X-SHOOTER instrument \citep{2006SPIE.6269E..98D,2011A&A...536A.105V} mounted at UT2 of the ESO-VLT on the nights of the 19th and 20th of July 2010. X-SHOOTER is a 3-arm single object spectrograph covering an extremely wide spectral range from 300 to 2500 nm. The spectral resolution of the instrument varies from  $R\sim3300$ to $R=5500$ depending on wavelength for a 1.2$\arcsec$ slit.\\
The observations took place just after a cold front had passed over Paranal, and as a result the observing conditions were extremely poor, windy, non photometric and with poor seeing. The seeing was of the order of 1.5$\arcsec$ in average, and up to 2$\arcsec$. The total time that could be spent on the targets ended up being much lower than originally planned, and different slits were used in a desperate effort to catch up with the seeing conditions. The detailed observation parameters are provided in Table \ref{tab:obs_details}. The three objects were observed three times with dithering between exposures. A limiting magnitude of  $\sim$23.5 was achieved at the wavelength of the \nb~by measuring the standard deviation over 5 \AA~and over the whole width of flux calibrated spectra. \\
The data were reduced using the EsoRex X-SHOOTER pipeline version 1.3.7, including sky subtraction and removal of bad pixels and cosmic rays. The calibrated spectra were then stacked using different methods: average, median and average with rejection of outliers. The following analysis has been carried out using the combination of the average with rejection of outliers spectra for each object.

\begin{table*}\centering
 \begin{tabular}{ccccccc}
 \hline\hline
Id       & $\alpha (J2000)$& $\delta (J2000)$ &\multicolumn{3}{c}{Exposure time (s)}                                                                                 & slit size \\
\hline 
          &                          &                             &  UVB                                           & VIS                                        & NIR                                 &\\
\hline
Obj1   &  334.143        & -17.5566                &3$\times$1700                          & 3$\times$1700                    & 3$\times$1800              & 1.2''  \\
Obj2   & 334.099         & -17.5326                & 2$\times$1100 + 600               & 2$\times$1100 + 600          & 3$\times$1200              & 1.2''\\ 
Obj3  &   334.208        & -17.5349                & 3$\times$1750 + 1287              & 3$\times$1750 + 1287        & 3$\times$1800              & 1.5''\\

\hline
\end{tabular}
\caption{List of the slits and exposure times for each of three X-SHOOTER arms used for each of the three objects. Dithering was applied between individual exposures.}
\label{tab:obs_details}
\end{table*}

\begin{figure*}\centering
    \includegraphics[scale=0.55]{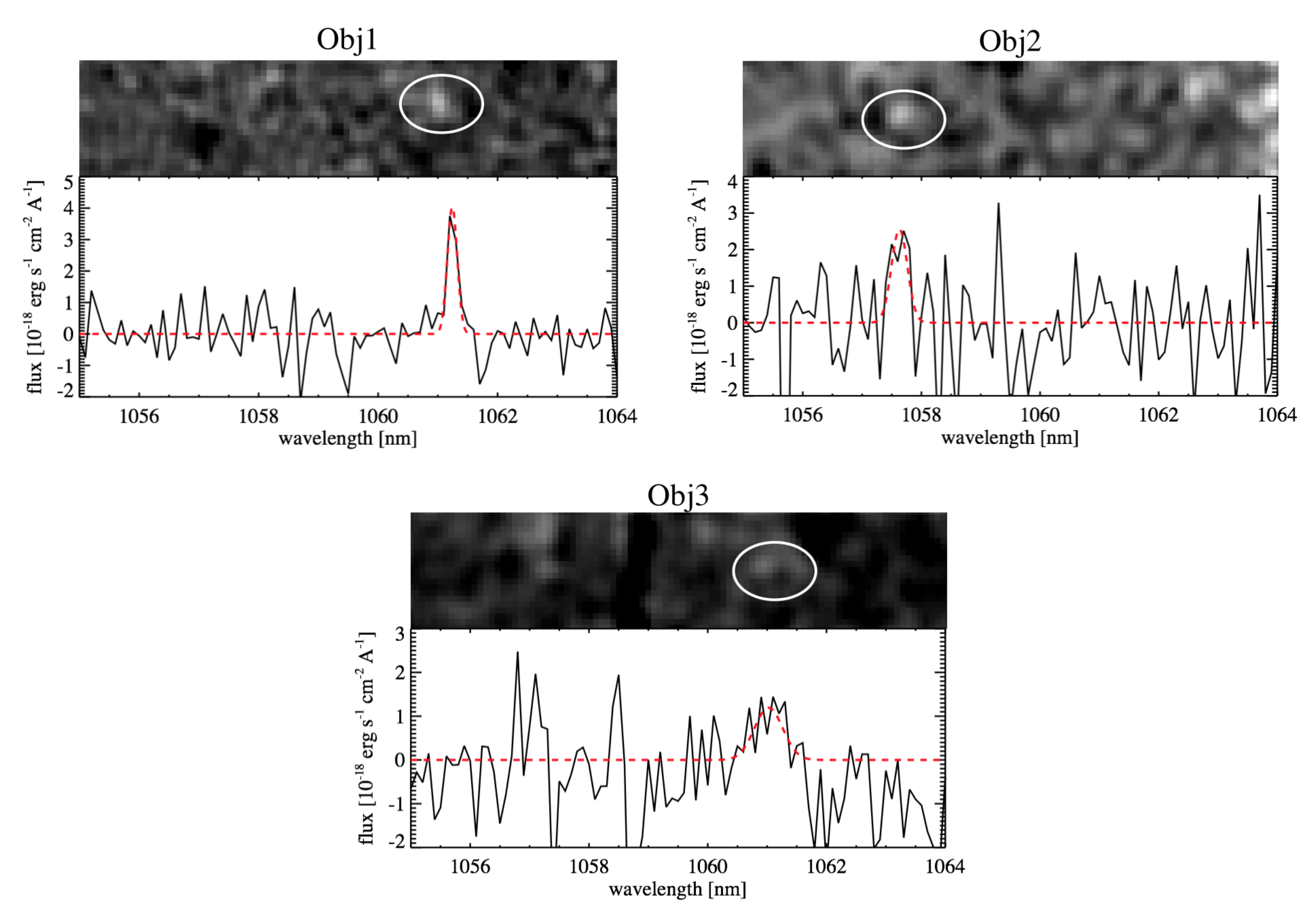}
  \caption{2D and 1D spectra centered on the line detected in \nb~ for each object. The red dashed line is the Gaussian fit of the line.}
  \label{fig:lines+gauss}
\end{figure*}

\begin{figure*}\centering
    \includegraphics[scale=0.75]{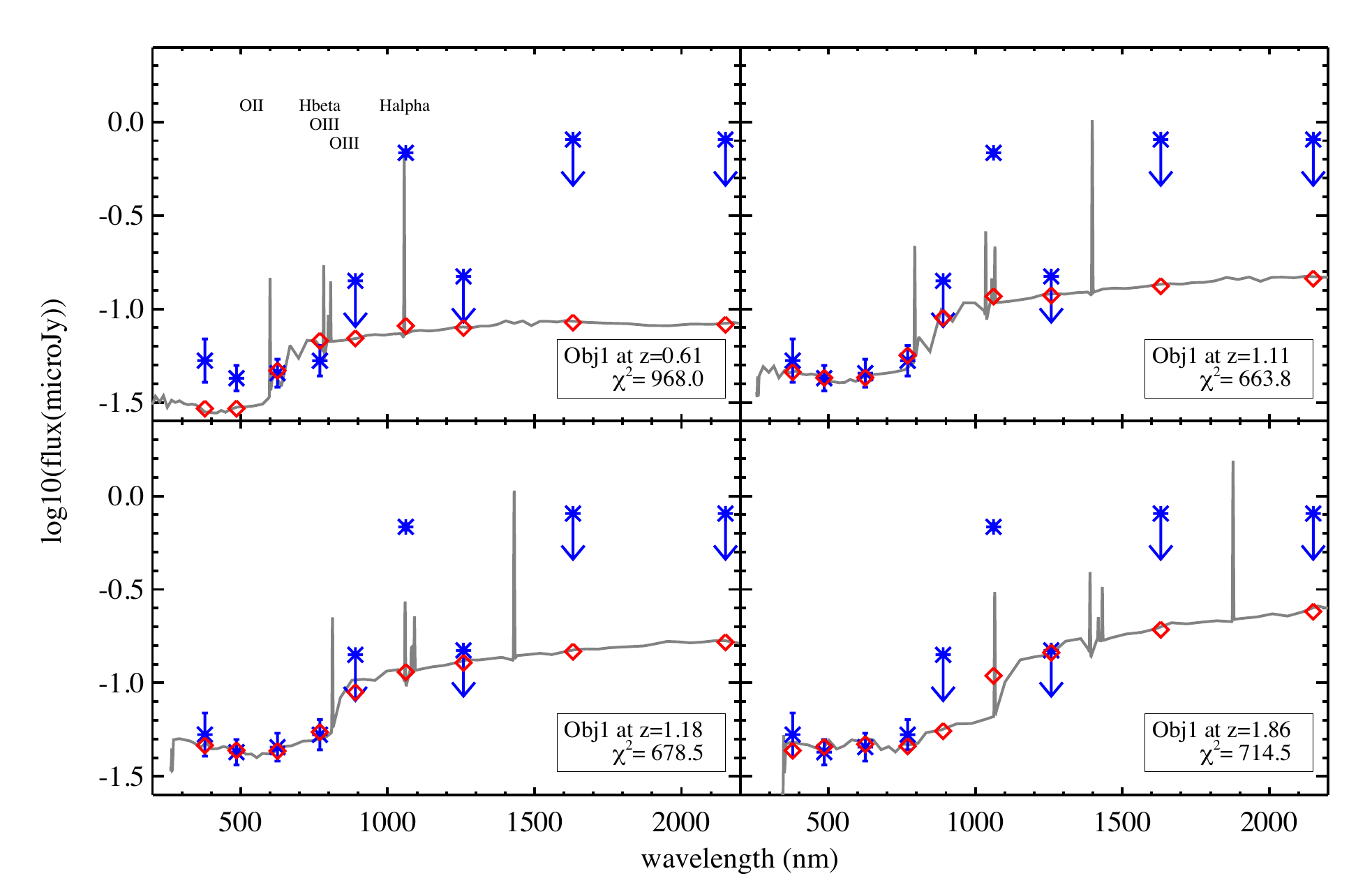}
  \caption{Results of the SED fitting with LePhare for Obj1 using \citet{2003MNRAS.344.1000B} models and assuming that the detected line is H$\alpha$ at $z\sim$0.61, O[III] at $z\sim1.11$, H$\beta$ at $z=1.18$, and O[II] at $z\sim1.86$. The grey line is the best fit SED. Blue crosses are measured magnitudes or magnitude upper limits while red squares are the expected magnitudes coming from the fit.}
  \label{fig:spec_all_lp}
\end{figure*}


\section{Analysis}\label{par:which_obj}
Emission lines were searched in each of the 2D stacked spectra from a careful visual inspection. Interestingly, we only detect one emission line in each of the three 2D spectra within the  \nb~ filter bandwidth. The reality of these lines was carefully checked by tracking their spatial positions along the slit on the 2D spectrum on each of the three individual dithered images forming the 2D stacks. For each object we extract a 1D line spectrum. Both 1D and 2D spectra for each of the three objects are shown in Fig.~\ref{fig:lines+gauss}.\\
For each object, we evaluated the slit losses when measuring the line fluxes. These slit losses were evaluated considering the (poor) seeing conditions prevailaing during the observations, the slit width, and the sizes of each object as measured on the NB images. The accuracy of the estimate of the slit losses is low, and is estimated to be of the order of 30\%. This introduces systematic uncertainties in the scientific analysis that follows, without however altering our main conclusions.\\
The level of the noise was estimated on the 1D spectra by measuring the standard deviation on both sides of the line on $\sim$30 pixels (equivalent to $\sim$3\AA).
\subsection{Methodology and application to Object 1}
The central wavelength of the line is $\lambda=$1061.3 nm. Fitting a Gaussian profile to the line, and after applying a correcting factor of 2 to account for slit losses, we measure an integrated flux of 1.50$\times10^{-17}$ erg s$^{-1}$ cm$^{-2}$ with a SNR of 7.7.\\
The measured spectroscopic line flux is equivalent to an  magnitude of 24.55 over the bandwith of the narrow band filter, indicating that the line flux indeed dominates the \nb~magnitude of the object (24.3). We derive a continuum level of $\sim 3.93\times10^{-20}$ erg s$^{-1}$ cm$^{-2}\AA^{-1}$, corresponding to a magnitude of 26.0. This continuum level is far too faint to be detected on the XSHOOTER data, and is also consistent with the non detection in the J-band at the 5$\sigma$ limit of our observations, (AB$\sim$26). From this continuum level, we also derive an EW of the line of $\sim 400$ \AA. We note that the absolute flux measurements are associated to significant uncertainties (due to the errors on slit losses for instance), leading in turn to large uncertainties on the estimate of the EW.\\
The detection of a single line prevents a straightforward identification of the redshift of the source. For the purpose of determining this redshift from the whole photometric and spectroscopic dataset, we resort to a SED fitting analysis using LePhare \citep{1999MNRAS.310..540A,2006A&A...457..841I}. This code uses a $\chi^2$ template fitting method, and for the purpose of our analysis we use spectral templates of dusty starburst galaxies with emission lines.\\
We first used templates from \citet{2003MNRAS.344.1000B}, subsequently modified by LePhare with variable extinction laws  \citep{1994ApJ...429..582C} and addition of emission lines following classical recipes that relate galaxy star formation rates and luminosities in the UV continuum, recombination lines and forbidden lines \citep{1998ARA&A..36..189K}. Moreover, LePhare was run by constraining the redshift to discrete intervals for which one of the O[II], O[III]4959 \AA, O[III]5007 \AA, H$\beta$, and H$\alpha$ lines corresponds to the detected emission line. Best fits are shown on Fig. \ref{fig:spec_all_lp} with their associated $\chi^2$ values. Overall, no good fit including the NB flux was found at any redshifts by LePhare by changing the extinction and the emission lines. \\
We therefore investigate extreme spectra with high EW emission lines. We use spectra of Ultra Strong Emission Line object \citep[USEL,][]{2007ApJ...668..853K,2009ApJ...698.2014H}. USELs are characterised by strong emission lines that can have extreme restframe equivalent widths (EWs) of several hundreds \AA. We therefore use a representative USEL spectrum to evaluate whether their spectral properties fit our data. We first run LePhare, similarly to the starburst case. The results of the fits and of the $\chi^2$ values are shown in panel (a) of Fig. \ref{fig:fit_usel_all}. The used USEL spectrum, representative of the USEL population, provides a much better fit to the observed photometric data of the object than a dusty starburst spectrum coming from the library of \citet{2003MNRAS.344.1000B} if the detected line is O[III](5007) \AA. Indeed, the reduced $\chi^2\sim1$ as compared to several hundreds.\\
We then investigate whether other emission lines should have been detected in the XSHOOTER data. Here again, we investigate two types of spectra:  standard star-forming galaxies with and without extinction \citep{2011A&A...525A.143C} and USELs. We assume that the detected line is one of the five O[II], O[III]4959 \AA, O[III]5007 \AA, H$\beta$, and H$\alpha$ emission lines, and we compute in turn the signal to noise ratio expected in our data on the three other lines. The line ratios without extinction are indicated in Table~\ref{tab:flux_ratios}. In the case of star forming galaxies without extinction, it is impossible to reproduce a situation where only one line could be detected. In all situations, at least one line should have been detected with a signal to noise ratio above 2. However, considering a relatively high level of extinction of $E(B-V) = 0.4$, the single detection of the H$\alpha$ line in the NB filter is possible. Nevertheless, this possibility has been excluded by the LePhare analysis. In the case of USELs, if the measured line is O[III]5007 \AA, no other lines could have been detected in our data (signal to noise ratios between 1 and 2). Indeed, the O[III]5007 \AA line strongly dominates the USEL spectra (see Table~\ref{tab:flux_ratios}). No coincidence with sky absorption or emission lines could have explained the single line detection in our data in all other situations than this line being O[III]5007 \AA.\\
In conclusion of this section, there is convincing evidence that Obj1 is a USEL O[III] emitter at $z=1.12$.

\subsection{Object 2}
The central wavelength of the line is $\lambda = 1057$ nm. Fitting a Gaussian profile and applying the slit loss correction factor of 2, its integrated flux is $1.71 \times10^{-17}$ erg s$^{-1}$ cm$^{-2}$ with a SNR of 5.1. Over the bandwidth of the \nb~filter the measured spectroscopic flux is equivalent to an  magnitude of 24.4, therefore the line flux dominates the \nb~magnitude of the object (24.3). The derived continuum level is $\sim1.8 \times10^{-20}$ erg s$^{-1}$ cm$^{-2}\AA^{-1}$ corresponding to a magnitude of 26.8 which is again consistent with the non-detection in the J-band at 5$\sigma$. Using that level of continuum we derive an EW of $\sim1000\AA$.\\
LePhare does not allow to find a suitable fit solution with \citet{2003MNRAS.344.1000B} templates whereas the USEL template provides a good fit to the observed photometric data as shown in panel (b) of Fig. \ref{fig:fit_usel_all} (reduced $\chi^2\sim2$ as compared to several hundreds). The check for the detection of other emission lines leads to either the strong O[III]5007\AA line of a USEL or the H$\alpha$ line of an extinct star-forming galaxy, the latter being inconsistent with the LePhare results.\\
In conclusion, Obj2 is also a USEL O[III] emitter at $z=1.12$ and the EW of the O[III] line from the USEL template is 830 \AA, consistent, within error bars, with our observations.


\subsection{Object 3} 
\begin{figure*}\centering
    \includegraphics[scale=0.6]{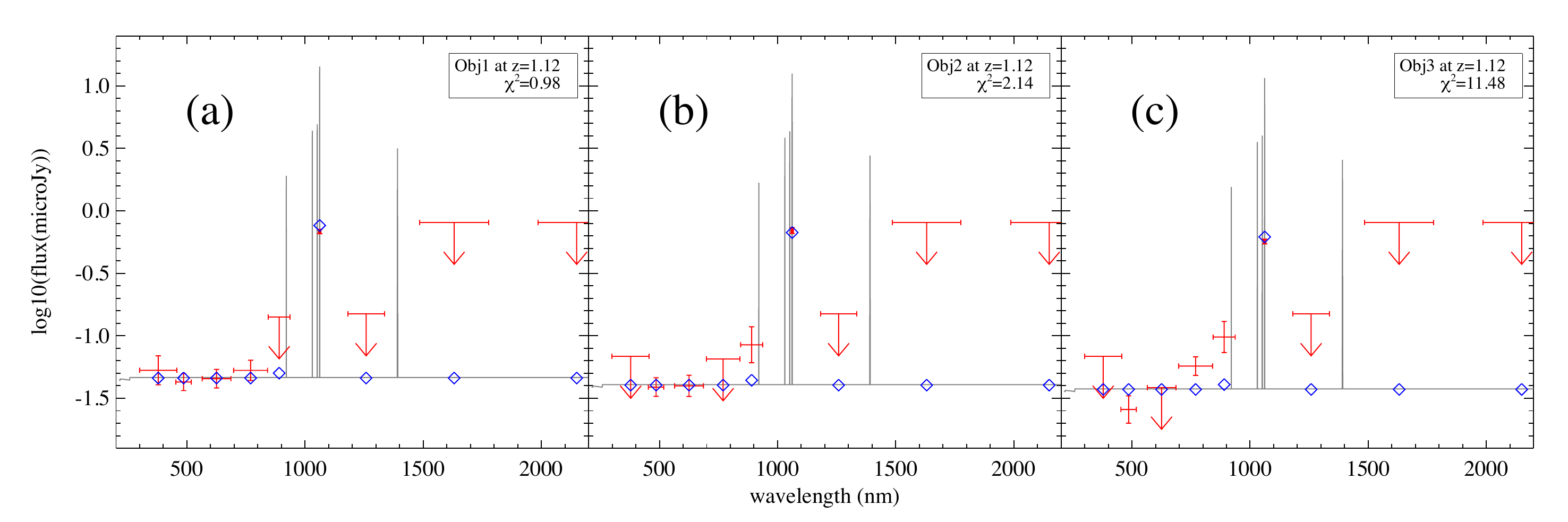}
  \caption{Fits of the three objects magnitudes using LePhare with a USEL template assuming that the detected line is O[III]5007 \AA. Red crosses are the measured magnitudes while arrows are upper limits. The blue diamonds show the magnitudes predicted by LePhare and the grey line is the best fit spectrum from LePhare. }
  \label{fig:fit_usel_all}
\end{figure*}

One single emission line is detected in the 2D spectrum with a central wavelength of $\lambda = 1061$ nm (see Fig. \ref{fig:lines+gauss}). The fit of a Gaussian profile to the line and the application of a correcting factor of 1.7 to account for slit loss lead to a measured integrated flux of $1.27 \times10^{-17}$ erg s$^{-1}$ cm$^{-2}$ with a SNR of 3.2. This line flux is equivalent to a magnitude of 24.7 over the bandwidth of the \nb~filter, the \nb~magnitude is therefore dominated by the line flux of the object (24.52). The estimated  continuum level is $\sim3.0 \times10^{-20}$ erg s$^{-1}$ cm$^{-2}\AA^{-1}$ corresponding to a magnitude of 26.3. Similarly to Obj1, the continuum is too faint to be detected in the XSHOOTER data and is consistent with the non-detection in the J-band (AB$\sim$26 at 5$\sigma$). The EW of the line associated with this continuum is $\sim420\AA$.\\
We carry the same analysis using LePhare as for Obj1. No good fit solution was found from the templates of  \citet{2003MNRAS.344.1000B} (reduced $\chi^2$ of several hundreds). We checked if other emission lines could have been detected in both cases, standard star-forming galaxies (with and without extinction) and USELs. In the former case, the detected line could be H$\alpha$ as the signal to noise ratio of the other lines are lower than 2$\sigma$. However this possibility is ruled out by the LePhare analysis. Indeed, the magnitude in the \nb~filter predicted by the template fitting is strongly inconsistent with the observed magnitude (2 mag difference). Considering USELs, only the detection of the O[III] line is consistent with no detection of other lines in our data. However the LePhare fit is not satisfying either as the reduced $\chi^2$ is $\sim11$, as shown in panel (c) of Fig. \ref{fig:fit_usel_all}. \\
In conclusion, Obj3 is likely to be a USEL O[III] emitter at $z=1.12$. The O[III] EW measured from the USEL template is $\sim$670 \AA which is consistent, within the error bars, with our observations.

\begin{table}\centering
 \begin{tabular}{lcc}
 \hline\hline
line ratio                               & Star-forming              & USELs \\
\hline 
O[III]5007/O[III]4959           & 3                                 & 3\\
$\mathrm{O[III]5007}$/O[II]                 & 0.35                            & 4.5\\     
H$\alpha$/O[II]                   & 1                                & 2.5\\
H$\beta$/H$\alpha$          & 0.35                            & 0.35\\
\hline
\end{tabular}
\caption{The two sets of flux ratios considered in that study. Those for star-forming galaxies come from \citet{2011A&A...525A.143C} and those for USELs are from \citet{2007ApJ...668..853K} and \citet{2009ApJ...698.2014H}.}
\label{tab:flux_ratios}
\end{table}

\section{Discussion and implication for high-$z$ surveys}\label{par:disc}
We followed spectroscopically three peculiar objects selected as high-EW line emitters in a search for LAE candidates at $z\sim7.7$ using very deep optical data \citep{2012A&A...538A..66C}. Had we used optical data shallower by approximately one magnitude, we would have selected these three objects as genuine LAE candidates at $z\sim7.7$. These objects have magnitude differences between the five optical (ugriz) bands and the \nb~band ranging from 2.3 to 3.5. This represents large color breaks, considering that the optical - NB breaks used for selecting LAE samples at high redshift is of the order of 1.5 to 2.5 magnitudes in most studies \citep[for instance,][]{2012ApJ...744...89H,2012ApJ...752..114S,2014ApJ...797...16K}. 
Indeed, from Table 1, we infer that the use of color breaks one magnitude lower the three objects would have been selected as LAE candidates. Furthermore, these objects could also contaminate high-$z$ LBG samples through broad band detection of their strong line emission. In a thorough study of LBG selection criteria,  \citet{2011ApJ...730...68C} remarked that USELs are a potential source of interlopers in $z > 7$ galaxy searches and conclude that the $i$-band data must be 4 magnitudes deeper than the $J$-band data. \\
These three objects are therefore potential contaminants to high-$z$ galaxy samples and unveiling their nature is of interest for estimating the contamination rate of these samples. Despite the low quality of our data due to poor observing conditions during the observations, we identify emission lines in spectroscopy at the position corresponding to the wavelength range of the \nb\ filter. For none of these objects do we find other emission lines in the XSHOOTER data. After SED fitting and detailed analysis of the spectroscopic data, we argue that the emission line in the \nb\ filter is O[III]5007\AA, placing these three contaminants at a redshift of $\sim$1.1.\\
This claim is consistent with similar reports of spectroscopically confirmed contaminants available in the literature that highlight the importance of O[III] emission. Such objects are also refered to as Extreme Emission Line Galaxies \citep[for instance]{2014arXiv1412.7909H}. In searching LBGs at redshift between 8 and 10 from deep near IR imaging, \citet{2003A&A...412L..57R} identified a O[III] emitter at z = 1.68 from spectroscopic follow-up observations). This object is characterised by a high EW and it has O[III]5007/H$\beta$ $\sim$5.9 fully consistent with USEL values. Similarly, \citet{2012MNRAS.425L..19H} followed a $z\sim11$ LBG candidate with XSHOOTER and detected several emission lines, amongst which O[III] is the strongest with an EW of $\sim$700 \AA, which put that galaxy at $z=2.08$. They conjecture that this object is either a heavily obscured starburst or an old galaxy upon which a burst of star formation is superimposed. The example of UDFj-39546284 is illustrative of the extreme difficulty of confirming very high-redshift galaxy candidates and of the role that faint low-$z$ interlopers may have. This object was originally claimed to be at $z\sim10$ \citep{2011Natur.469..504B}. With deeper near IR data, \citet{2013ApJ...763L...7E} suggested that this object could lie at an even higher redshift of $z=11.9$. From HST WFC3 grism observations \citet{2013ApJ...765L...2B} detected a 2.7$\sigma$ line that could either be Ly-$\alpha$ at $z=12.12$ or O[III] at $z=2.19$, \citet{2013ApJ...765L..16B} finding the later solution more plausible. From Keck-MOSFIRE data,  \citet{2013ApJ...773L..14C} could not improve the line detection level and therefore conclude on the nature of this object. All these examples suggest the prevalence of O[III] emitters amongst interlopers to high-$z$ galaxy candidates, in line with our findings.\\
Finally, as a last sanity check, we estimate the number of strong O[III] emitters expected in the CFHTLS D4 field from their Luminosity Function at redshift 0.8 as reported in Fig. 13 of \citet{2007ApJ...668..853K}. We assume no evolution in redshift between 0.8 and 1.1, and we further assume that the \nb\ flux is entirely dominated by O[III] emission. The redshift interval probed by the \nb~filter is 0.02 for the O[III] line, corresponding in turn to a comoving volume of $\sim3\times10^{3}$ Mpc$^3$ for our CFHTLS D4 field. We consider a \nb~limiting magnitude of 24.6, corresponding to a line luminosity of 10$^{41}$ erg.s$^{-1}$. We derive that there should be of the order of four ultra-strong O[III] emitters in our data at the flux limit that we consider, in good agreement, within low number statistics and large uncertainties, with the three objects that we selected. It is re-assuring that both numbers match, comforting the likelihood that the three objects are indeed O[III] emitters. Note that there were seven objects in the CFHTLS-D4 field displaying similar photometric properties, the other four objects have \nb~magnitudes fainter than 24.6.\\
We infer from this study that O[III] emitting objects are a likely important source of contamination in NB-selected high-redshift galaxy samples, and preventing contamination from these objects require optical data deeper than the NB data by almost 3 magnitudes. As noted above, \citet{2011ApJ...730...68C} suggest 4 magnitudes difference to prevent $z>7$ galaxy (LBG) samples from being contaminated by low-$z$ interlopers.\\
Interestingly, a majority of the $z\sim 2.2$ H$\alpha$ emitter candidates originally selected from NB imaging at 2.2 $\mu$m in \citet{2000A&A...362....9M} turned out to be O[III] emitters after detailed follow-up analyses \citep{2003mglh.conf..302M}. \citet{2014MNRAS.440.2375M} and \citet{kochia} recently performed photometric redshift analyzes of NB emitters at 1.18 $\mu$m and  1.06 $\mu$m respectively, and they find a fraction of H$\beta$/O[III] emitters of 40\% and 28\% respectively. Combined to an analysis of the broad band photometric data, these analyses could inform us further on the likelihood of strong O[III] emitters in NB studies.\\
The occurrence of strong O[III] emitters seems to be ubiquitous over a large redshift range, and such objects are a likely source of contamination/error in many studies. A more detailed analysis of the O[III] luminosity functions at redshifts up to 3 is needed to further evaluate the level of contamination that O[III] might generate in several of the planned dark energy surveys. Strong O[III] emission might contaminate photometric redshifts, be they directly used for scientific analyses (e.g. Euclid) or for photometric selection of targets to be subsequently followed-up in spectroscopy (e.g. PFS/SuMIRe or DESI). We suggest that the O[III] luminosity function should be better determined over a large redshift range, as recently completed by \citet{2014arXiv1408.1523C} for the O[II] luminosity function over the redshift range [0.1 - 1.65]. Such a study can be carried out using the WISP (WFC3 Infrared Spectroscopic Parallel) survey \citep[PI:Malkan]{2010ApJ...723..104A} in which several hundreds of USELS have been found by \citet{2014ApJ...789...96A} over the redshift range $0.3<z<2.3$. Even if it is crucial, the characterisation of the O[III] luminosity function at higher redshift will be possible only with IR space telescopes such as the JWST. Only the detailed knowledge of the O[III] luminosity function could allow us to better estimate the contamination rate of any high-z galaxy survey from O[III] emitters.

\section*{Acknowledgment}
The authors would like to thank the referee whose comments improved this manuscript. We thank Hakim Atek, Brian Lemaux, Danka Paraficz, Johan Richard, and Tayyaba Zafar for valuable discussions. We thank Len Cowie for providing the USEL templates. \\
This work received support from the Agence Nationale de la Recherche bearing the reference ANR-09-BLAN-0234. JPK acknowledges support from ERC advanced LIDA and from CNRS.

\bibliographystyle{aa}

\bibliography{biblio}
\end{document}